\documentclass[graybox,envcountchap]{svmult}

\usepackage[english]{babel}
\usepackage{amsmath}
\usepackage{amssymb}
\usepackage{bm}
\usepackage{chapterbib}
\usepackage{mathtools}
\usepackage{multicol}
\usepackage{multirow}
\usepackage{rotating}
\renewcommand{\Re}{\mathop\mathrm{Re}\nolimits}

\newcounter{chapGiaquinta}

\makeatletter
\newcommand{\manuallabel}[2]{\def\@currentlabel{#2}\label{#1}}
\makeatother

\newcommand{\sortrelax}[1]{{\relax}} 

\usepackage{mathptmx}        
\usepackage{helvet}          
\usepackage{courier}         

\usepackage{makeidx}         
\usepackage{graphicx}        
\usepackage{multicol}        
\usepackage[bottom]{footmisc}

\usepackage{trfsigns} 
\usepackage{mhchem} 

\usepackage{wasysym} 

\usepackage{tabularx} 

\newcommand\clearrow{\global\let\rowmac\relax}
\clearrow

\makeatletter
\def\printindex#1#2{\@restonecoltrue\if@twocolumn\@restonecolfalse\fi
  \columnseprule \z@ \columnsep 35pt
  \csname phantomsection\endcsname
  \chapter*{#2}
  \markright{\uppercase{#2}}
  \addcontentsline{toc}{chapter}{#2}
  \begin{multicols}{2}
  \@input{#1.ind}%
  \end{multicols}%
}
\makeatother



\begin{document}


    \thispagestyle{empty}
    \par\ \par
    \vspace*{2in}
     \begin{flushleft}
     {\large\bfseries Many-body approaches at different scales} \\[\baselineskip]
     {\large A tribute to Norman H. March on the occasion of his 90th birthday} \\[29pt]
     G. G. N. Angilella and C. Amovilli, \emph{Editors} \\[29pt]
     (New York : Springer, 2017)\\
     ISBN 978-3-319-72373-0 (hardback)\\
     ISBN 978-3-319-72374-7 (ebook)\\
     DOI 10.1007/978-3-319-72374-7
    \end{flushleft}
    \vfill
    \pagebreak\par

\frontmatter

\advance\textheight by -8\baselineskip



\advance\textheight by 8\baselineskip


\mainmatter





\setcounter{chapter}{11}
\title{Kovacs effect and the relation between glasses and supercooled liquids\protect\setcounter{chapGiaquinta}{\thechapter}}
\manuallabel{chap:Giaquinta}{\thechapGiaquinta}

\author{F. Aliotta \and R. C. Ponterio \and F. Saija \and P. V. Giaquinta}

\index{author}{Aliotta, F.}
\index{author}{Ponterio, R. C.}
\index{author}{Saija, F.}
\index{author}{Giaquinta, P. V.} 

\institute{%
F. Aliotta \and R. C. Ponterio \and F. Saija
\at
CNR-IPCF, Viale F. Stagno d'Alcontres 37, 98158 Messina, Italy
\at \email{aliotta@ipcf.cnr.it} , \email{ponterio@ipcf.cnr.it}, \email{franz.saija@cnr.it}
\and
P. V. Giaquinta
\at Dipartimento di Scienze Matematiche e Informatiche, Scienze Fisiche e Scienze della Terra,
Universit\`a degli Studi di Messina,
Contrada Papardo, I-98166 Messina, Italy
\at \email{paolo.giaquinta@unime.it}
}

\maketitle

\abstract{%
In this note we revisit the Kovacs effect, concerning the way in which the volume of a glass-forming liquid, 
which has been driven out of equilibrium, changes with time while the system evolves towards a metastable state. 
The theoretical explanation of this phenomenon has attracted much interest even in recent years, because of its 
relation with some subtle aspects of the still elusive nature of the glass transition. 
In fact, even if there is a rather general consensus on the fact that what is experimentally observed on 
cooling is the dramatic effect produced by the dynamical arrest of slower degrees of freedom over the experimental time scale, 
it is not yet clear whether this phenomenology can be justified upon assuming the existence of an underlying 
(possibly, high order) phase transition at lower temperatures.
}

\index{keyword}{liquids}
\index{keyword}{liquids!supercooled}
\index{keyword}{supercooled liquids}
\index{keyword}{Kovacs effect}
\index{keyword}{ortho-terphenyl}

\section{Introduction}

Understanding the kinetic and thermodynamic routes followed by a supercooled liquid while becoming, 
through viscous slowdown, a glass well below the freezing point still represents a major challenge 
in the chemical physics of condensed matter \cite{Biroli:13,Ediger:12}. In fact, in spite of countless 
efforts on both the theoretical and experimental sides, many significant questions on the very nature 
of the glassy state still remain unanswered. Even if there is a widespread consensus on the thesis 
that the experimentally observed glass transition is the macroscopic outcome of relaxation processes 
which, at low enough temperatures, become much slower than the experimental observation time, 
it is not clear yet whether such a slowdown of the dynamics of the system can be interpreted, 
following Kauzmann's original discussion on this point \cite{Kauzmann:48}, as an indication of an 
underlying, possibly continuous, phase transition, which would occur below the vitrification point $T_{\textrm{g}}$. 
In addition, it is not clear whether what appears -- on the experimental time scale -- as a kinetically 
arrested system may eventually transform into a truly metastable state after a sufficiently long time. 
However, distinguishing between broken ergodicity and metastability is not an easy task on the operational 
side and both laboratory and numerical experiments give ambiguous indications on this 
point~\cite{Pusey:86,Lubchenko:07,Stillinger:88,Eckmann:08,vanMegen:98,Boue:11,Mallamace:14}. 

Moreover, other relevant questions remain open, one of which concerns the surmised existence of a 
liquid-liquid phase transition which would be undergone by deeply supercooled metastable 
water \cite{Debenedetti:03,Stanley:07,Speedy:82,Mishima:85,Stanley:97,Sastry:96} and the very possibility 
of prolonging the associated coexistence line well below the homogeneous nucleation temperature, 
so as to verify whether it eventually merges with the coexistence line between the experimentally 
observed low-density and high-density amorphous phases of water~\cite{Mishima:98,Mishima:02}.

The focus of this note is on a phenomenon, intimately associated with the glass transition, 
that was originally observed and described by Kovacs~\cite{Kovacs:64}. This phenomenon gives 
useful information on the volumetric time evolution of a glass-forming liquid, which has 
been originally driven out of equilibrium. The experimental protocol implemented by 
Kovacs entails three stages: (i) a thermodynamically stable liquid, formerly equilibrated 
at a temperature $T_{\textrm{i}}$, is quenched, at a fixed pressure, to a temperature 
$T_{\textrm{q}}$, not too far below $T_{\textrm{g}}$, in such a way that some internal 
degrees of freedom fall out of equilibrium with the thermal bath; (ii) the system is 
then left to age for some time, which, however, is not long enough for it to reach a 
condition of full thermal equilibrium; (iii) the temperature is finally raised to a 
value $T_{\textrm{f}}$, intermediate between $T_{\textrm{i}}$ and $T_{\textrm{q}}$. 
One can then observe the irreversible evolution of the system as it relaxes from 
the preset out-of-equilibrium condition at $T=T_{\textrm{q}}$ to an asymptotic 
one of metastable equilibrium at $T=T_{\textrm{f}}$. In particular, the way 
in which the volume of the sample changes with time exhibits a ``memory'' effect in 
that it is found to depend in a non trivial way on the thermal history of the material. 
In fact, if the final temperature is not too high, the system starts expanding irreversibly 
to a volume which, after overshooting the equilibrium value at $T_{\textrm{f}}$, 
further increases up to a maximum value that is lower (and reached later) 
the higher the quenching temperature $T_{\textrm{q}}$. Thereafter, the system 
progressively contracts until it regains its equilibrium volume at $T=T_{\textrm{f}}$~\cite{Mossa:04}. 
The nonmonotic behaviour of the volume and the resulting maximum imply that the values of pressure, 
temperature, and volume are not sufficient to identify a unique (nonequilibrium) state of the material: 
in fact, under the conditions outlined above, for assigned values of such three variables, 
the system can be actually observed in two different ``states'', corresponding to different 
stages of the dynamical evolution of the material towards metastable equilibrium.

The Kovacs effect, originally observed in a polymeric substance (polyvinyl acetate), 
has been observed in a variety of glassy materials. As such, it has been the topic of many experimental 
and theoretical investigations. In more recent times Angell and coworkers \cite{Angell:00} discussed the 
phenomenon in the framework of the volumetric behaviour of glass formers in nonergodic regimes, 
with specific reference to nonlinear relaxation and associated memory effects. 
Mossa and Sciortino~\cite{Mossa:04} performed molecular dynamics simulations of a model of ortho-terphenyl 
(OTP) which revealed some fine details of the dynamics of the phenomenon that are not accessible 
to laboratory experiments; they also performed an analysis of the properties of the potential 
energy landscapes explored by the system during the relaxation process. A theoretical interpretation 
of these results was later attempted by Bouchbinder and Langer~\cite{Bouchbinder:10}, who resorted to a 
description of the system based on (separable) configurational and kinetic-vibrational subsystems.

The aim of this note is to illustrate a simple macroscopic model which can be used to 
describe the dynamic evolution and, correspondingly, the thermodynamic behaviour of a 
system along the lines traced by Kovacs with its experimental protocol. We shall also 
discuss the implications of the proposed model as to some aspects of the thermodynamic 
relation between glasses and metastable liquids.

\section{Kovacs effect in ortho-terphenyl}

As is well known, the glass transition does not occur at a well defined temperature but, 
rather, over a range of temperatures across which, depending on the time scale of the experiment, 
a number of internal degrees of freedom of the system become \emph{de facto} arrested. 
An indirect measure of the effective number of energetically active degrees of freedom 
over the experimental time scale is given by the isobaric specific heat which, contextually, 
exhibits a rather sharp drop across $T_{\textrm{g}}$. 
Correspondingly, cusp discontinuities show up in the thermal behaviour of the extensive parameters. 
However, no latent heat is released which implies that the entropy is continuous across the glass transition.

\begin{figure}[th]
\centering
\includegraphics[width=0.8\textwidth]{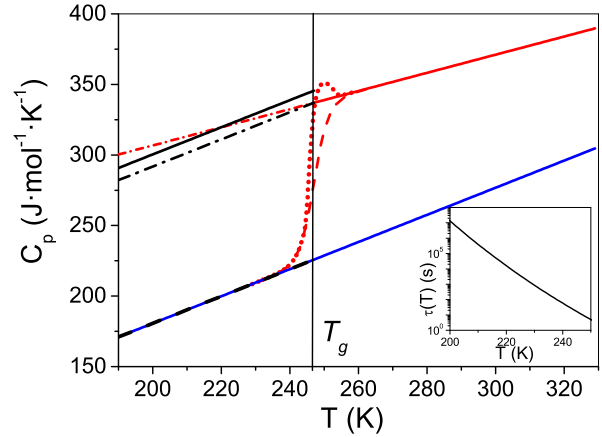}
\caption{Isobaric molar specific heat of ortho-terphenyl plotted plotted as a function of the 
temperature below the freezing/melting point at typical modulation angular frequencies 
($\approx 10^{-1}$~Hz). 
Continuous red line: supercooled-liquid branch (linear fit of the experimental data \cite{Chang:72}); 
dash-dotted red line: linear extrapolation of the supercooled-liquid data below the glass transition point; 
continuous blue line: solid branch (linear fit of the experimental data \cite{Chang:72}); 
dashed red line: crossover between the liquid and glassy branches modelled through Eq.~\ref{eq:Giaquinta:1} 
and using the relaxation time for slow processes displayed in the inset (see text); 
dotted red line: typical crossover between the glassy and liquid branches on fast 
heating the partially aged glass; dash-dotted black line: specific heat of the glass 
obtained after quenching the metastable fluid at $T=T_{\textrm{g}}$; 
continuous black line: effective specific heat of the quenched and partially 
aged glass with an associated fictive temperature $\widetilde{T}_{\textrm{g}} < T_{\textrm{g}}$ (see text). 
The vertical black line marks the glass transition point ($T_{\textrm{g}}=247$~K).}
\label{fig:Giaquinta:1}
\end{figure}

\begin{figure}[th]
\centering
\includegraphics[bb=43 21 457 566,clip,width=0.6\columnwidth]{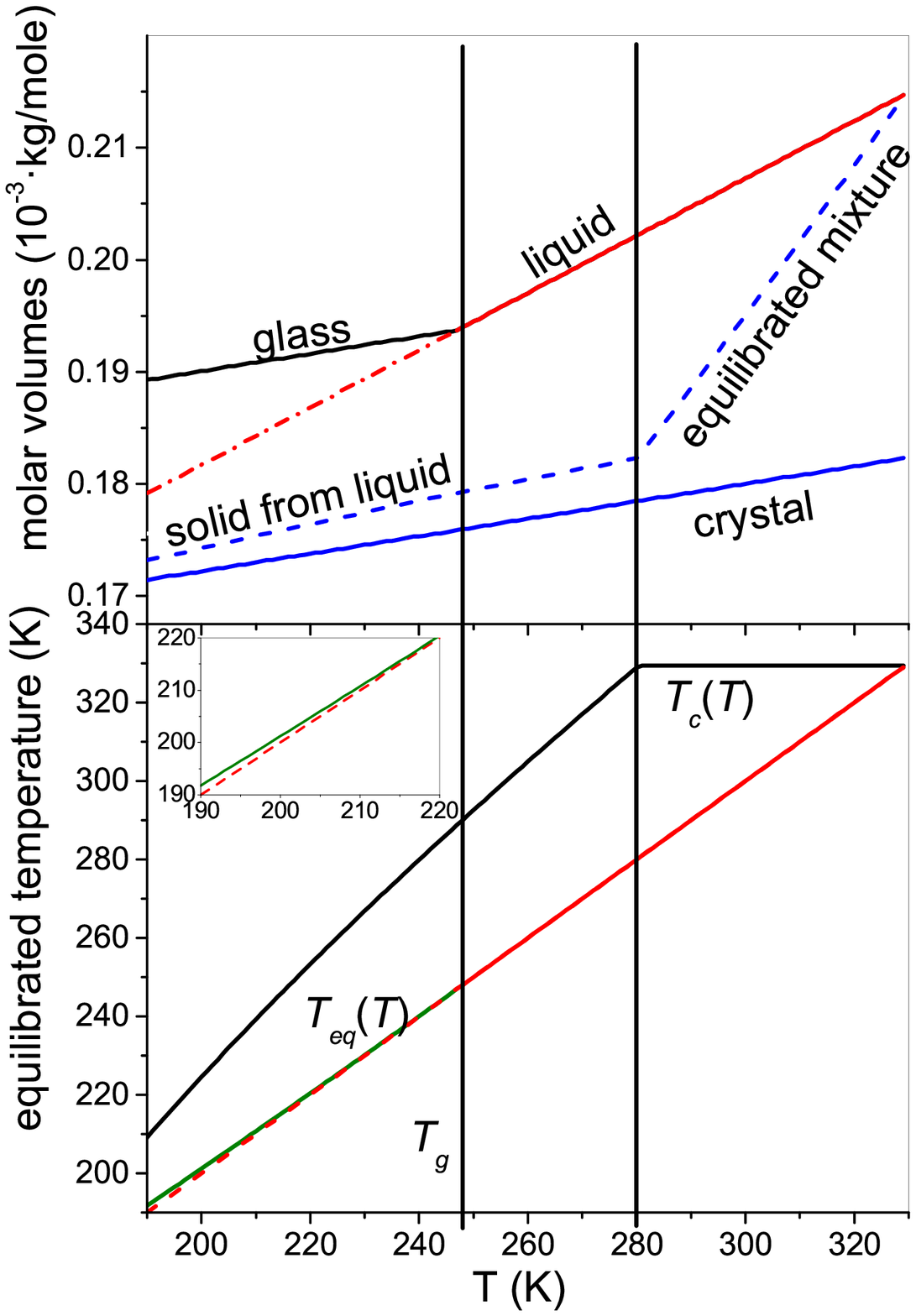}
\caption{Upper panel: molar volume of ortho-terphenyl plotted as a function of the temperature below 
the freezing point at ambient pressure. Red continuous line: supercooled liquid; 
red dash-dotted line: linear extrapolation of the supercooled-liquid data below the glass transition point; 
black continuous line: glass; blue dashed line: molar volume of the stable phase nucleated by metastable liquid 
OTP upon spontaneous freezing at adiabatic-isobaric (\emph{i.e.,} isoenthalpic) conditions. 
The cusp singularity at $T_{\textrm{solid}}\approx 280$~K marks the boundary between two different outcomes 
of the irreversible transition eventually undergone by supercooled liquid OTP: for $T>T_{\textrm{solid}}$ 
the nucleated phase is a solid-liquid mixture at the freezing/melting temperature $T_{\textrm{m}}$, 
whereas for lower temperatures the equilibrium phase is a pure crystalline solid whose temperature 
decreases with (while still keeping higher than) that of the parent liquid, and whose volume is 
correspondingly larger than that of the solid (blue continuous line) at $T_{\textrm{m}}$ 
(for more details see \cite{Aliotta:13}). Lower panel: temperature of the asymptotic 
metastable phase to which the liquid, originally undercooled to a temperature $T$, would 
relax at isoenthalpic conditions (black continuous line); for graphical convenience, 
we also plot the temperature of the metastable supercooled liquid as a red line -- 
the dashed part being the linear extrapolation below $T_{\textrm{g}}$ of the higher-temperature experimental data -- 
which, by construction, coincides with the first quadrant bisector; the green continuous line 
(also expanded in the inset) represents the temperature $T_{\textrm{eq}}$ of the metastable state 
that would be reached asymptotically by the glass as calculated through Eq.~(\ref{eq:Giaquinta:4}) 
in the text. The two black vertical lines mark the temperatures $T_{\textrm{g}}$ and $T_{\textrm{solid}}$, respectively.}
\label{fig:Giaquinta:2}
\end{figure}

In the following we shall use OTP as a reference material for our discussion of the Kovacs effect. 
Liquid OTP can be easily supercooled below its freezing/melting temperature ($T_{\textrm{m}} = 329.35$~K), 
down to the glass transition point that is located at a relatively high temperature ($T_{\textrm{g}} \approx 247$~K). 
In addition, the glass can be slowly reheated and restored to the metastable liquid phase; 
this is possible because the homogeneous nucleation temperature ($T_{\textrm{H}}$) of this material is 
lower than $T_{\textrm{g}}$. Chang and Bestul used adiabatic calorimetry to measure the heat capacities of 
liquid, glassy and crystalline OTP at ambient pressure \cite{Chang:72}. These data are plotted in Fig.~\ref{fig:Giaquinta:1}. 
The heat capacities of both liquid and crystalline OTP are found to be nearly linear functions of the temperature. 
Moreover, the heat capacity of glassy OTP  ($T < T_{\textrm{g}}$) is only slightly higher ($1.5\%-2\%$) than that of the crystal, 
a difference that is not resolved on the scale of the plot displayed in Fig.~\ref{fig:Giaquinta:1}. 
As noted above, the glass transition is signalled by the abrupt drop of the specific heat; correspondingly, 
a cusp discontinuity shows up in the molar volume of OTP \cite{Naoki:89}, that is plotted in the upper panel of Fig.~\ref{fig:Giaquinta:2}.

When a system, which was formerly at thermodynamic equilibrium, is cooled, it will start relaxing 
towards a new equilibrium condition, a process which implies a redistribution of the internal 
energy among all the degrees of freedom as well as a variety of local and global structural 
rearrangements over distances and times which can be very long. If the time window of 
the experimental observation is fixed, only the motions which take place over shorter times 
will be able to relax. Instead, slower motions will appear somewhat frozen in a state corresponding 
to that of the system at equilibrium at the initial temperature. Of course, the location of a temporal 
``boundary'' between frozen and active motions depends on the experimental time window. 
This is the reason why the experimental glass transition temperature cannot be defined in an unambiguous way.

In order to take explicitly into account this crucial aspect, we assumed that the relaxation of 
the temperature-dependent molar specific heat of our model system, as observed in a typical differential 
scanning calorimetry experiment, can be represented by the expression \cite{Schawe:95,Alig:97}:
\begin{equation}
C_P(T;\omega)=\Re \left[C_P^{\textrm{(c)}}(T) + \frac{C_P^{\textrm{(l)}}(T) - C_P^{\textrm{(c)}}(T)}{1 + j\omega\tau(T)}\right],
\label{eq:Giaquinta:1}
\end{equation}
where $C_P^{\textrm{(l)}}(T)$ and $C_P^{\textrm{(c)}}(T)$ are the isobaric specific heats of the 
liquid and of the crystal, respectively, $\tau(T)$ is the (temperature dependent) relaxation 
time of the one single slow process which characterizes the kinetics of the model, 
$\omega=2\pi t{_{\textrm{p}}^{-1}}$ is the reverse of the experimental sampling 
time $t_{\textrm{p}}$, and $j$ is the imaginary unit. The inverse Fourier transformation of Eq.~(\ref{eq:Giaquinta:1}) 
leads to the following time dependence of the isobaric specific heat at a given temperature $T$ \cite{Schawe:95}: 
\begin{equation}
C_P(T;t)=C_P^{\textrm{(c)}}(T) + \left[ C_P^{\textrm{(l)}}(T) - C_P^{\textrm{(c)}}(T) \right]\left[1-e^{-\frac{t}{\tau(T)}}\right].
\label{eq:Giaquinta:2}
\end{equation}
The liquid and crystal specific heats show up in the above equations as the long-time (zero-frequency) 
and short-time (infinite-frequency) values of $C_P(T)$, respectively.

To be more realistic one should actually consider a distribution of relaxation times, 
whose widths would also depend on the temperature, as well as a more plausible model for 
the complex susceptibility of the system. However, here we are not as much interested in 
reproducing the experimental data for OTP in a detailed and quantitative way; in fact, 
we just want to show that the effect originally observed by Kovacs is the natural outcome 
of a relaxation process occurring in the system, even in the oversimplified case in 
which this process is being parametrized with one single relaxation time only. 
As for its dependence on the temperature, we adopted an Arrhenius-like expression: 
$\tau(T) = \tau_{0} \exp (\Delta E/k_{B}T)$, 
and adjusted the values of the two free parameters so as to obtain a r
ough match with the experimental values reported for OTP in \cite{Mallamace:14}. 
We also assumed that the relaxation times of the fast processes are much shorter 
than the observation time in the experiment under consideration. The resulting behaviour of 
the specific heat across the glass transition region is displayed for a typical angular frequency in 
Fig.~\ref{fig:Giaquinta:1}, whose inset shows the relaxation time that we plugged into Eq.~(\ref{eq:Giaquinta:1}). 

As already noted, when a system, which has previously attained thermodynamic equilibrium 
at a temperature very close to $T_{\textrm{g}}$, is rapidly cooled to a lower temperature, 
fast motions rapidly equilibrate in the new thermal state, whereas slow motions remain substantially 
``frozen'' in the configurational state that the system was in at $T=T_{\textrm{g}}$. 
Hence, we can estimate the ``effective'' specific heat of the quenched fluid as:
\begin{equation}
{C}_P^{\textrm{(eff)}}(T) \approx C_P^{\textrm{(c)}}(T) + \left[C_P^{\textrm{(l)}}(T_{\textrm{g}}) - C_P^{\textrm{(c)}}(T_{\textrm{g}})\right] ,
\label{eq:Giaquinta:3}
\end{equation}
where the term in square brackets on the r.h.s. of Eq.~(\ref{eq:Giaquinta:3}) is the jump observed in the specific heat of 
the system at the glass transition point, which approximately quantifies the ``hidden'' contribution to 
${C}_P^{\textrm{(eff)}}(T)$ that is not resolved by calorimetric measurements on the time scale of the experiment. 
The statement embodied in Eq.~(\ref{eq:Giaquinta:3}) conveys an information analogous to that 
which emerges from the thermal behaviour of the molar volume of the vitrified system; 
in fact, the experimental data for $v_{\textrm{g}}(T)$ (see Fig.~\ref{fig:Giaquinta:2}) can be reproduced rather accurately by the expression:
\begin{equation}
v_{\textrm{g}}(T) \approx v_\textrm{c}(T) + \left[v_\textrm{l}(T_{\textrm{g}}) - v_\textrm{c}(T_{\textrm{g}})\right] ,
\label{eq:Giaquinta:4}
\end{equation}
where $v_\textrm{l}$, $v_{g}$ and $v_\textrm{c}$ are the molar volumes of supercooled liquid, 
glass, and crystal, respectively. Hence, as already noted 
before \cite{Nieuwenhuizen:01}, a system which has undergone a dynamical arrest at $T_{\textrm{g}}$ 
and which has been then quenched to a lower temperature $T_{\textrm{q}}$ can be characterized by 
two temperatures: the actual quenching temperature and an auxiliary temperature ($T_{\textrm{g}}$) 
at which the slow motions have \emph{de facto} arrested over the experimental time scale. 
Yet, as time goes on even such slow configurational degrees of freedom start to relax, 
gradually driving the system towards a condition of metastable equilibrium. As far as 
specific heat and volume are concerned, such an asymptotic state will correspond to a 
point on the lines traced upon extrapolating (with constant slope, at least for moderate
 amounts of supercooling) the laboratory data of the liquid branch (see Figs.~\ref{fig:Giaquinta:1} and \ref{fig:Giaquinta:2}). 
 Under adiabatic conditions, this process towards equilibrium will necessarily imply a transfer 
 of energy from the fast motions to the slower ones and, correspondingly, a change of volume. 
 According to this description, the final temperature ($T_{\textrm{eq}}$) will fall between 
 the glass transition temperature and the quenching temperature. In other words, the 
 out-of-equilibrium glass, obtained through the rapid quenching of the system, 
 will irreversibly relax towards a condition of metastable equilibrium characterized by a higher temperature, 
 and -- at least, in the case of OTP -- a higher density. We can calculate such intermediate temperature 
 by noting that, under the postulated adiabatic conditions, the enthalpies of the initial and final ``states'' 
 should be equal, \emph{i.e.,} $H^{\textrm{(eff)}}(T_{\textrm{q}})=H^{\textrm{(l)}}(T_{\textrm{eq}})$; hence, it follows that:
\begin{equation}
\int_{T_{\textrm{q}}}^{T_{\textrm{m}}}{C_P^{\textrm{(eff)}}(T) dT} = \int_{T_{\textrm{eq}}}^{T_{\textrm{m}}}{C_P^{\textrm{(l)}}(T) dT},
\label{eq:Giaquinta:5}
\end{equation}
where the melting/freezing temperature $T_{\textrm{m}}$ has been assumed as a common 
reference temperature for evaluating the enthalpy changes of the system along two paths starting 
from the unrelaxed and relaxed state, respectively. Obviously, the resulting final equilibrium 
temperature $T_{\textrm{eq}}$ is a function of the temperature $T_{\textrm{q}}$ at which the 
system had been previously quenched. As seen in the lower panel of Fig.~\ref{fig:Giaquinta:2}, 
the system undergoes a moderate heating upon irreversibly relaxing to a metastable condition.

We shall now explore what happens if the system is allowed to exchange energy with 
an external thermostatic reservoir, which brings us into the conditions of the experiment performed by Kovacs. 
We consider it useful to carry out our conceptual experiment using two different protocols, the second 
of which complies more closely with Kovacs' indications.

\begin{figure}[th]
\centering
\includegraphics[width=0.8\columnwidth]{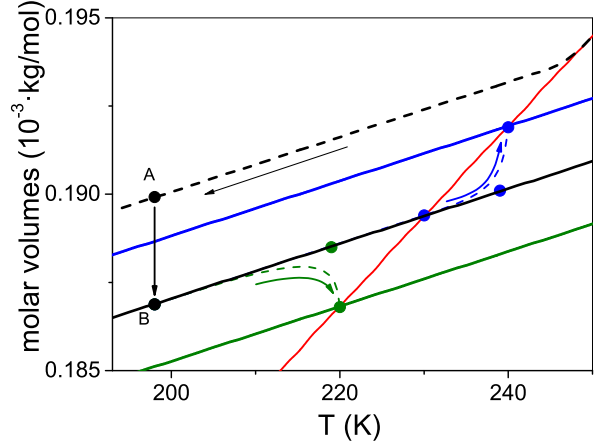}
\caption{Modified Kovacs' protocol (see text): the sample is equilibrated at $T_{\textrm{g}}$ 
and then rapidly cooled down to 198.5~K: the continuous red line is the supercooled liquid line 
(linearly extrapolated below $T_{\textrm{g}}$); the dashed black line represents the molar 
volume under the above condition and the solid black circle labelled A marks 
the molar volume attained by the system after quenching; the solid black circle 
labelled B represents the molar volume achieved by the partially aged system. 
The continuous black line represents the glass branch that would be followed 
by the system with a fictive temperature corresponding to the intersection 
point with the liquid branch. Experiment \#1: the sample at B is rapidly heated to $240$~K; 
the blue circles and the blue arrow indicate the time evolution of the molar volume. 
The blue continuous line represents the glass branch with fictive temperature $\widetilde{T} = 240$~K. 
Experiment \#2: the sample at B is rapidly heated to $220$~K; the green circles and the green arrow 
indicate the time evolution of the molar volume. The green continuous line represents the glass 
branch with fictive temperature $\widetilde{T} = 220$~K.}
\label{fig:Giaquinta:3}
\end{figure}

We first assume that liquid OTP has been equilibrated down to the 
ordinary vitrification threshold. We then imagine to cool rapidly the system from $T_{\textrm{g}} = 247$~K to a lower temperature, 
say $T_{\textrm{q}}=198.5\,\textrm{K}$. As a result, the system contracts and its molar 
volume decreases to the value $v^\textrm{(A)}_{\textrm{g}}$, corresponding to point A on the line $v_{\textrm{g}}(T)$ 
displayed in Fig.~\ref{fig:Giaquinta:3}. Let the system now exchange energy with the thermal reservoir at 
the quenching temperature for a time $(t_{\textrm{aging}})$ long enough for its configurational 
state to change, but nevertheless shorter for the system to reach equilibrium. In order to keep 
temperatures within a range compatible with the available experimental data for OTP, 
we chose $t_{\textrm{aging}}=10^7$~s. While relaxing at constant pressure, the system contracts further. 
Correspondingly, the isobaric specific heat changes according to Eq.~(\ref{eq:Giaquinta:2}). 
After the prescribed time has elapsed, the isobaric specific heat, calculated through Eq.~(\ref{eq:Giaquinta:2}), 
has attained the (larger) value $C_{P}(T_{\textrm{q}};t_{\textrm{aging}})=295.8\,\textrm{J}\,\textrm{mol}^{-1}\,\textrm{K}^{-1}$. 
Following the same line of thought illustrated before on discussing Eq.~(\ref{eq:Giaquinta:3}), 
we can infer the ``fictive'' temperature $(\widetilde{T}_{\textrm{g}})$ at which the system 
would have effectively deviated from the metastable-liquid branch, had it been cooled at a 
\emph{slower} rate than that leading to vitrification at $247$~K. Coherently with the assumption underlying Eq.~(\ref{eq:Giaquinta:3}), we write:
\begin{equation}
{C_P}(T_{\textrm{q}};t_{\textrm{aging}}) \approx C_P^{\textrm{(c)}}(T_{\textrm{q}}) + \left[C_P^{\textrm{(l)}}(\widetilde{T}_{\textrm{g}}) - C_P^{\textrm{(c)}}(\widetilde{T}_{\textrm{g}})\right].
\label{eq:Giaquinta:6}
\end{equation}
Equation~(\ref{eq:Giaquinta:6}) allows us to determine the fictive temperature 
($\widetilde{T}_{\textrm{g}}\approx 230$~K) at which the liquid-to-solid jump of the 
isobaric specific heat is such that the value derived from Eq.~(\ref{eq:Giaquinta:6}) is equal to that provided by Eq.~(\ref{eq:Giaquinta:2}) 
for $T=T_{\textrm{q}}$ and $t=t_{\textrm{aging}}$. Note that, for temperatures lower than $T_{\textrm{g}}$, 
the specific heat of the metastable liquid used in Eq.~(\ref{eq:Giaquinta:6}) was estimated through a 
linear extrapolation of the experimental data. The state of this partially relaxed glass -- whose volume, 
after quenching and aging, has so far dropped to the value $v^\textrm{(B)}_{\textrm{g}}$ (corresponding to point B in Fig.~\ref{fig:Giaquinta:3}) 
-- is equivalent to that which would be produced upon rapidly cooling a liquid whose slow dynamics has arrested at $230$~K 
(instead than at $247$~K). This latter statement is crucial for explaining the outcome of Kovacs' experiment, 
whose third and final stage consists in heating the system to a temperature $T_{\textrm{f}}$, intermediate between $T_{\textrm{q}}$ 
and $T_{\textrm{g}}$. In fact, we imagine that, as soon as the time $t= t_{\textrm{aging}}$ has elapsed, 
the system is immediately coupled with another thermostat whose temperature is $T_{\textrm{f}}$ and 
then left free to relax until (metastable) equilibrium has been eventually reached.

We shall now investigate the dynamical behaviour of the system, with specific regard to the way 
in which the volume changes with time, when partially aged glassy OTP is heated to different temperatures 
from similarly prepared samples (\emph{i.e.,} samples quenched to the same temperature and aged for the same time). 
In this way, the outcome of the heating procedure will not be influenced by differing initial conditions of the material. 
In fact, the states of, say, equally aged samples are generally different at different quenching temperatures because 
the relaxation times of the system depend on the temperature.

We start inspecting the volumetric behaviour of the system when the final temperature is \emph{higher} 
than the fictive temperature $\widetilde{T}_{\textrm{g}}$, while being lower than $T_{\textrm{g}}$ (see Fig.~\ref{fig:Giaquinta:3}). 
Let us choose $T_{\textrm{f}}=240$~K. In the short-time regime, only the fast degrees of freedom of the system 
react to the modified thermal condition; hence, the volume starts increasing from the value $v^\textrm{(B)}_{\textrm{g}}$, 
closely following, as the system warms up, the glass line $v_{\textrm{g}}(T;\widetilde{T}_{\textrm{g}})$ 
that intercepts the supercooled-liquid branch at $T=\widetilde{T}_{\textrm{g}}$. 
As soon as the system approaches and eventually surpasses the vitrification threshold, 
the slower configurational degrees of freedom start relaxing as well. As a result, 
the volume keeps growing \emph{monotonically}, while departing from the glass line, 
until the system has eventually equilibrated at the prescribed temperature. 

Let us now set the final temperature, at which the system -- 
previously prepared in the same initial state B as in the thought experiment discussed above -- 
is to be heated, to a value \emph{lower} than $\widetilde{T}_{\textrm{g}}$, say $T_{\textrm{f}}=220$~K: in such conditions 
(see Fig.~\ref{fig:Giaquinta:3}), the molar volume of the partially-aged glass at $T_{\textrm{f}}$ 
turns out to be larger than the molar volume of the metastable liquid (extrapolated) at $T=220$~K. Hence, even in 
this case we would again observe an expansion of the system at short times: the volume would initially 
increase following the glass line which departs from the metastable branch at the fictive temperature calculated above. 
However, as soon as the slow configurational degrees of freedom become active, the molar volume would start 
shrinking, after the initial rise, so as to approach the lower value which corresponds to the asymptotic 
equilibrium state at the prescribed final temperature.

Hence, in this second thought experiment the molar volume will exhibit a \emph{non-monotonic} time behaviour. 
The resulting maximum is the distinguishing feature that was originally observed by Kovacs \cite{Kovacs:64}. 
In the present scheme, as is manifest from Fig.~\ref{fig:Giaquinta:3}, on approaching the equilibrium value 
the volume passes through a maximum only if the final temperature at which the system is heated is lower 
than the fictive temperature $\widetilde{T}_{\textrm{g}}$. Moreover, the difference between 
the maximum value attained by the molar volume and the asymptotic equilibrium value turns out to be larger 
the larger the difference $(\widetilde{T}_{\textrm{g}}-T_{\textrm{f}})$.

The experimental protocol discussed above partially differs from that originally 
designed by Kovacs. In fact, this author reported on the time behaviour of the 
volume of a system which was heated to one single temperature $T_{\textrm{f}}\le T_{\textrm{g}}$ from 
several lower temperatures at which the system had been previously cooled from the same equilibrium 
state at a temperature $T_{\textrm{i}}\ge T_{\textrm{g}}$. Before heating the system, the quenched 
liquid was left to age for a variable timespan: the lower the quenching temperature was, the longer the aging time would be. 

In order to ``simulate'' Kovacs' protocol, we proceed as before, assuming that the system has been cooled 
from $T_{\textrm{g}}$ to a temperature $T_{\textrm{q}}$ in the range $190\,\textrm{K} - 210$~K. The just formed 
glass is then allowed to relax but for not so long that it may reach equilibrium, say for a time of the order 
of $10^7$~s. Correspondingly, the volume of the aged glass decreases from the value attained soon after 
quenching, whereas the specific heat increases. Using Eq.~(\ref{eq:Giaquinta:2}), we calculate the value of 
the specific heat pertaining to this new state, \emph{i.e.,} ${C_P}(T_{\textrm{q}};t_{\textrm{aging}})$, which, 
through Eq.~(\ref{eq:Giaquinta:6}), allows us to infer the fictive temperature $\widetilde{T}_{\textrm{g}}$. 
Once we know this datum, we can calculate the value of the molar volume of the partially aged glass which, 
in our picture, will be equal to the molar volume at $T=T_{\textrm{q}}$ of a glass whose dynamics 
has arrested at $T=\widetilde{T}_{\textrm{g}}$.

We now imagine to put the system in contact with a thermostat at the (higher) 
temperature $T_{\textrm{f}} = 230$~K; such a re-heating cycle is repeated a number of times, 
with the same target temperature but starting from different quenching temperatures in the cited range. 
In a rather short time (let us say, largely overestimating it, $10$~s), fast motions will have fully 
relaxed whereas the slow degrees of freedom are still frozen. As soon as fast relaxations have occurred, 
both the specific heat and the molar volume of the glass have contextually increased to the values which 
correspond to the higher temperature $T=T_{\textrm{f}}$ on the glass line, departing from the 
metastable liquid branch at $T=\widetilde{T}_{\textrm{g}}$. We now use Eq.~(\ref{eq:Giaquinta:6}), 
where $C_P^{\textrm{(c)}}$ has been substituted with ${C_P}(T_{\textrm{q}};t_{\textrm{aging}})$, 
to estimate the value attained by the specific heat once the system has been left free to relax for 
other $10$~s. Following the same procedure outlined above, we then calculate the new fictive 
temperature corresponding to the updated value of the specific heat at $t=20$~s, and the resulting 
value of the molar volume. Upon iterating this procedure, we can trace, with steps of $10$~s each, 
the time evolution of the molar volume of glassy OTP when it has been heated to $T_{\textrm{f}}$ from different quenching temperatures.

\begin{figure}[th]
\centering
\includegraphics[width=0.8\columnwidth]{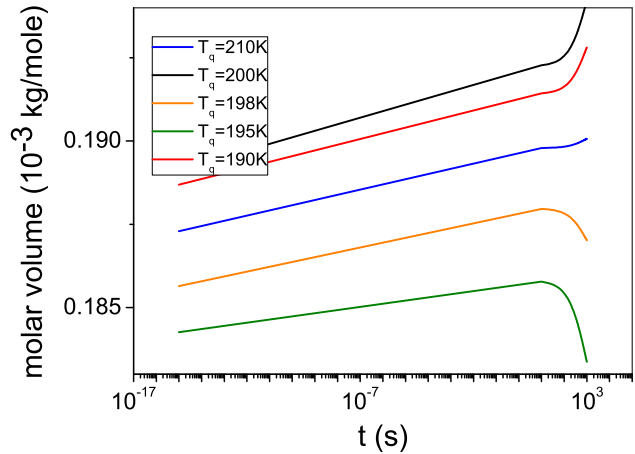}
\caption{Time evolution of the molar volume according to Kovacs' protocol: the system equilibrated 
at $T_\textrm{g}$ is rapidly cooled to a temperature $T_\textrm{q}$ in the range $190\,\textrm{K} - 210\,\textrm{K}$ 
and then aged for $10^{7} \,\textrm{s}$. After aging, the system is rapidly heated to 230~K and then left free to relax to thermal equilibrium.}
\label{fig:Giaquinta:4}
\end{figure}

The results predicted by our simplified model are shown in Fig.~\ref{fig:Giaquinta:4} and show that, 
coherently with what has been observed in both real and numerical experiments \cite{Kovacs:64,Mossa:04}, 
the molar volume of quenched OTP reaches its asymptotic equilibrium value $v_{\textrm{l}}(T_{\textrm{f}})$ 
either rising monotonically from the value of the partially aged glass or passing through a maximum at 
intermediate times, after having initially overshooted $v_{\textrm{l}}(T_{\textrm{f}})$. 
In the present scheme the occurrence of one or the other alternative behaviour critically 
depends on whether the final temperature $T_{\textrm{f}}$ is higher or lower than the temperature 
$T_{\textrm{X}}$ at which the extrapolated supercooled-liquid line and the \emph{effective} glass line, 
onto which the representative state of the glass has ``shifted'' after the initial aging, cross each other. 
In both scenarios, the molar volume of the heated glass will initially jump to a value larger than its initial 
value after quenching and aging. However, if $T_{\textrm{f}}>T_{\textrm{X}}$ the specific heat of 
the rapidly heated glass initially overshoots the value corresponding to the metastable liquid at the same temperature; 
from there on, the specific heat will decrease with time and this implies a gradual increase of the fictive 
glass temperature that is calculated at each step and, contextually, of the molar volume of the system. 
On the other side, if $T_{\textrm{f}}<T_{\textrm{X}}$ the specific heat attains a value that is smaller 
than that corresponding to the metastable liquid at the same final temperature; hence, it will keep on 
growing with time, which implies that the fictive glass temperature decreases as also does the molar volume.

Following the above discussion we can also interpret other aspects of the glass phenomenology. 
Imagine, for instance, that we perform a differential scanning calorimetry measurement on a 
glass which has been previously aged at low temperatures, in such a way that the rate at which 
the temperature is being changed corresponds to the time scale over which slow motions relax at $T=T_{\textrm{g}}$. 
Because of the aging, slow motions are initially equilibrated at a fictive temperature lower than $T_{\textrm{g}}$. 
Hence, when the temperature approaches $T_{\textrm{g}}$, an excess of heat from the bath (with respect to the
 enthalpy reduction originally undergone by the material upon quenching) is required to fully equilibrate the system, 
thus producing the ``endothermic overshoot'' that is typically observed in the temperature evolution of the isobaric specific heat.

\section {Concluding remarks}

In this note we have revisited the Kovacs effect in a simplified picture of the relaxation 
dynamics underpinning the glass transition. In particular, we have assumed that the time 
dependence of the isobaric specific heat can be mimicked using just one relaxation time 
and that fast motions have fully and systematically relaxed over the time step of our calculations. 
Notwithstanding such rough approximations, the resulting model is found to reproduce correctly, 
on a qualitative basis, the main features of Kovacs' experiment and to convey some useful insight on the phenomenology of the glass transition. 

Our analysis is based on the premise that the glass obtained through the rapid cooling of 
a supercooled liquid is an out-of-equilibrium system which, however, may asymptotically evolve 
towards a metastable phase provided it is given enough time to relax. We assume that such a 
metastable phase cannot be distinguished from that of the ``parent'' supercooled liquid at a given temperature. 
The relaxation process undergone by the glass, named ``aging'', is associated with a gradual change of 
the molar volume and, correspondingly, of the structural configuration of the material. 
In this perspective, distinguishing a long-aged glass from a metastable liquid may reduce to a 
merely semantic question, with obvious consequences on the hypothesis of an underlying (thermodynamic) 
glass transition. In fact, the apparent differences between the two structural conditions of the material 
emerge at the crossover temperature at which the experimental time scale becomes shorter than the configurational relaxation time of the system.

However, the postulated equivalence between the asymptotically ``equilibrated'' glass 
and the corresponding metastable liquid may be disproved by the existence of a threshold below 
which the nucleation of the stable crystalline phase can no longer be avoided. In this respect, 
a candidate threshold would be the Kauzmann temperature \cite{Kauzmann:48} at which the entropies of 
the metastable liquid and of the thermodynamically stable crystalline solid become equal. However, 
it has been argued that the hypothetical coexistence of a liquid and a solid phase would not be 
possible at a temperature lower than the equilibrium coexistence temperature \cite{Hoffmann:12,Aliotta:13}. 
In fact, whenever a supercooled liquid escapes from metastability and freezes, it does so irreversibly 
and adiabatically with an increase of both entropy and temperature as a consequence of the release of heat. 
As a result, the transition towards stable equilibrium takes place exothermically and the system warms 
up while solidifying. Hence, it does not make much sense to compare the entropies of the two phases at the same temperature.

The spontaneous freezing of a a metastable liquid is also associated with a change of volume. 
The spontaneous formation of a finite solid embryo produces a relatively large density fluctuation 
which propagates at low frequency, while dissipating, across the whole sample \cite{Aliotta:14}. This also explains 
why metastable equilibrium is a robust structural condition against fluctuations: even when a 
thermodynamic fluctuation brings, locally, the system close to the boundary between the metastable 
and stable equilibrium basins in phase space, dissipative processes are able to back reflect 
the system trajectory towards the metastable basin.

On cooling, a crossover temperature can be eventually reached at which the volume of the metastable 
system is equal to the volume of the stable phase that is formed under adiabatic conditions \cite{Aliotta:13}. 
This is what happens to OTP, as indicated in the upper panel of Fig.~\ref{fig:Giaquinta:2}. At such a temperature, 
the structural re-arrangement which drives the system towards the stable configuration is just a local 
process which does not need to propagate in order to be completed. In such conditions, local fluctuations 
are not dissipated away and the local transition, which results in a local increase of the temperature, 
immediately produces a further fluctuation in the adjacent volumes which can propagate rapidly 
(over a time scale comparable with the time required for the local rearrangement of a few molecules) 
across the whole sample. This argument is a different way for saying that, at that temperature, the 
energetic barrier between metastable and stable equilibrium conditions likely disappears. In this perspective, 
the observation that in water the volume crossover takes place at a temperature that, at normal pressure, 
is very close to the widely accepted value of the homogeneous nucleation temperature may not be a mere coincidence \cite{Aliotta:13}. 
In water, this temperature is definitely higher than the experimental glass-transition temperature and this can explain 
why metastable liquid water cannot exist at temperatures close to $T_{\textrm{g}}$.

Following our argumentation, one would be led to deduce that the existence in water 
of a crossover temperature at which the transition between the metastable liquid phase and the stable 
crystalline phase becomes both adiabatic and isochoric strongly supports the idea that any observed 
amorphous phase observed at very low temperature, which may well appear stable over the observation time scale, 
has no thermodynamic counterpart and can only be described in kinetic terms. On the contrary, in 
the case of OTP the volume crossover occurs at a temperature lower than the experimental $T_{\textrm{g}}$. 
Such a difference in the behaviour of glass-forming liquids as far as the relation between $T_{\textrm{g}}$ 
and the homogeneous nucleation temperature is concerned, has been already noted several years ago \cite{Angell:81}.

Summing up, the nature of the glass obtained when a liquid is rapidly cooled to low temperatures, 
over times shorter than those required for a complete structural rearrangement of the system, 
depends on the relation existing between the state which has been produced and the metastability basin of the system. 
For moderate supercoolings, the achieved state is not disconnected from the basin of the metastable liquid phase: 
hence, the glass, while being out of equilibrium, might still evolve, in principle, towards metastability. 
However, when quenched at very low temperatures, the system may be driven to a state which is no 
longer accessible from a metastable disordered phase. In such conditions, should it be able to rearrange itself, 
its unique, ultimate fate would be that of transforming into a stable solid.

\begin{acknowledgement}
The authors belonging to the Institute for Chemical-Physical Processes (IPCF) of the National Research Council (CNR) 
recall with enthusiasm the visit that Professor N. H. March paid to their institute in 2010. 
PVG expresses his profound gratitude to Professor March who invited him to visit the Imperial College of Science and Technology in London (UK) 
and later, on repeated occasions, the Theoretical Chemistry Department of the University of Oxford (UK) 
in the earlier stages of his post-graduation career. Working with and learning from him has always been an influential, unforgettable experience.
\end{acknowledgement}

\bibliographystyle{spphys}
\bibliography{a,b,c,d,e,f,g,h,i,j,k,l,m,n,o,p,q,r,s,t,u,v,w,x,y,z,Angilella,zzproceedings,thisvolume}

\begin{thebibliography}{10}
\providecommand{\url}[1]{{#1}}
\providecommand{\urlprefix}{URL }
\expandafter\ifx\csname urlstyle\endcsname\relax
  \providecommand{\doi}[1]{DOI \discretionary{}{}{}#1}\else
  \providecommand{\doi}{DOI \discretionary{}{}{}\begingroup
  \urlstyle{rm}\Url}\fi

\bibitem{Biroli:13}
G.~Biroli, J.P. Garrahan, J. Chem. Phys. \textbf{138}(12), 12A301 (2013).
\newblock \newline\doi{10.1063/1.4795539}

\bibitem{Ediger:12}
M.D. Ediger, P.~Harrowell, J. Chem. Phys. \textbf{137}(8), 080901 (2012).
\newblock \newline\doi{10.1063/1.4747326}

\bibitem{Kauzmann:48}
W.~Kauzmann, Chem. Rev. \textbf{43}(2), 219 (1948).
\newblock \newline\doi{10.1021/cr60135a002}

\bibitem{Pusey:86}
P.N. Pusey, W.~van Megen, Nature \textbf{320}(6060), 340 (1986).
\newblock \newline\doi{10.1038/320340a0}

\bibitem{Lubchenko:07}
V.~Lubchenko, P.G. Wolynes, Ann. Rev. Phys. Chem. \textbf{58}(1), 235 (2007).
\newblock \newline\doi{10.1146/annurev.physchem.58.032806.104653}

\bibitem{Stillinger:88}
F.H. Stillinger, J. Chem. Phys. \textbf{88}(12), 7818 (1988).
\newblock \newline\doi{10.1063/1.454295}

\bibitem{Eckmann:08}
J.P. Eckmann, I.~Procaccia, Phys. Rev. E \textbf{78}, 011503 (2008).
\newblock \newline\doi{10.1103/PhysRevE.78.011503}

\bibitem{vanMegen:98}
W.~van Megen, T.C. Mortensen, S.R. Williams, J.~M\"uller, Phys. Rev. E \textbf{58}, 6073 (1998).
\newblock \newline\doi{10.1103/PhysRevE.58.6073}

\bibitem{Boue:11}
L.~Bou\'e, H.G.E. Hentschel, V.~Ilyin, I.~Procaccia, J. Phys. Chem. B \textbf{115}(48), 14301 (2011).
\newblock \newline\doi{10.1021/jp205773c}

\bibitem{Mallamace:14}
F.~Mallamace, C.~Corsaro, N.~Leone, V.~Villari, N.~Micali, S.~Chen, Sci. Rep. \textbf{4}, 3747 (2014).
\newblock \newline\doi{10.1038/srep03747}

\bibitem{Debenedetti:03}
P.G. Debenedetti, J. Phys.: Condens. Matter \textbf{15}(45), R1669 (2003).
\newblock \newline\doi{10.1088/0953-8984/15/45/R01}

\bibitem{Stanley:07}
H.E. Stanley, P.~Kumar, L.~Xu, Z.~Yan, M.G. Mazza, S.V. Buldyrev, S.~Chen, F.~Mallamace, Physica A \textbf{386}(2), 729 (2007).
\newblock \newline\doi{10.1016/j.physa.2007.07.044}

\bibitem{Speedy:82}
R.J. Speedy, J. Phys. Chem. \textbf{86}(6), 982 (1982).
\newblock \newline\doi{10.1021/j100395a030}

\bibitem{Mishima:85}
O.~Mishima, L.D. Calvert, E.~Whalley, Nature \textbf{314}(6006), 76 (1985).
\newblock \newline\doi{10.1038/314076a0}

\bibitem{Stanley:97}
H.E. Stanley, L.~Cruz, S.T. Harrington, P.H. Poole, S.~Sastry, F.~Sciortino,   F.W. Starr, R.~Zhang, Physica A \textbf{236}(1), 19 (1997).
\newblock \newline\doi{10.1016/S0378-4371(96)00429-3}

\bibitem{Sastry:96}
S.~Sastry, P.G. Debenedetti, F.~Sciortino, H.E. Stanley, Phys. Rev. E \textbf{53}, 6144 (1996).
\newblock \newline\doi{10.1103/PhysRevE.53.6144}

\bibitem{Mishima:98}
O.~Mishima, H.E. Stanley, Nature \textbf{392}(6672), 164 (1998).
\newblock \newline\doi{10.1038/32386}

\bibitem{Mishima:02}
O.~Mishima, Y.~Suzuki, Nature \textbf{419}(6907), 599 (2002).
\newblock \newline\doi{10.1038/nature01106}

\bibitem{Kovacs:64}
A.J. Kovacs, \emph{Transition vitreuse dans les polym{\`e}res amorphes. Etude ph{\'e}nom{\'e}nologique} (Springer Berlin Heidelberg, Berlin, Heidelberg, 1964), pp. 394--507.
\newblock \newline\isbn{978-3-540-37073-4}.
\newblock \newline\doi{10.1007/BFb0050366}

\bibitem{Mossa:04}
S.~Mossa, F.~Sciortino, Phys. Rev. Lett. \textbf{92}, 045504 (2004).
\newblock \newline\doi{10.1103/PhysRevLett.92.045504}

\bibitem{Angell:00}
C.A. Angell, K.L. Ngai, G.B. McKenna, P.F. McMillan, S.W. Martin, J. Appl. Phys. \textbf{88}(6), 3113 (2000).
\newblock \newline\doi{10.1063/1.1286035}

\bibitem{Bouchbinder:10}
E.~Bouchbinder, J.S. Langer, Soft Matter \textbf{6}, 3065 (2010).
\newblock \newline\doi{10.1039/C001388A}

\bibitem{Chang:72}
S.S. Chang, A.B. Bestul, J. Chem. Phys. \textbf{56}(1), 503 (1972).
\newblock \newline\doi{10.1063/1.1676895}

\bibitem{Aliotta:13}
F.~Aliotta, P.V. Giaquinta, M.~Pochylski, R.C. Ponterio, S.~Prestipino, F.~Saija, C.~Vasi, J. Chem. Phys. \textbf{138}(18), 184504 (2013).
\newblock \newline\doi{10.1063/1.4803659}

\bibitem{Naoki:89}
M.~Naoki, S.~Koeda, J. Phys. Chem. \textbf{93}(2), 948 (1989).
\newblock \newline\doi{10.1021/j100339a078}

\bibitem{Schawe:95}
J.E.K. Schawe, Thermochimica Acta \textbf{260}, 1 (1995).
\newblock \newline\doi{10.1016/0040-6031(95)90466-2}

\bibitem{Alig:97}
I.~Alig, Thermochimica Acta \textbf{304}, 35 (1997).
\newblock \newline\doi{10.1016/S0040-6031(97)00174-3}

\bibitem{Nieuwenhuizen:01}
T.M. Nieuwenhuizen, J. Chem. Phys. \textbf{115}(17), 8083 (2001).
\newblock \newline\doi{10.1063/1.1399036}

\bibitem{Hoffmann:12}
H.~Hoffmann, Mat.-wiss. u, Werkstofftech. \textbf{43}(6), 528 (2012).
\newblock \newline\doi{10.1002/mawe.201200673}

\bibitem{Aliotta:14}
F.~Aliotta, P.V. Giaquinta, R.C. Ponterio, S.~Prestipino, F.~Saija, G.~Salvato, C.~Vasi, Sci. Rep. \textbf{4}, 7230 (2014).
\newblock \newline\doi{10.1038/srep07230}

\bibitem{Angell:81}
C.A. Angell, E.J. Sare, J.~Donnella, D.R. MacFarlane, J. Phys. Chem. \textbf{85}(11), 1461 (1981).
\newblock \newline\doi{10.1021/j150611a001}

\end{thebibliography}











\end{document}